\documentclass{PoS}
\usepackage{amsmath}
\usepackage{braket}
\usepackage{bm}

\title{Neutrino eigenstates and flavour, spin and spin-flavour oscillations in a constant magnetic field}

\ShortTitle{Neutrino eigenstates and oscillations in a constant magnetic field}

\author{Alexey Lichkunov\\
        Department of Theoretical Physics, Moscow State University, 119992 Moscow, Russia\\
        E-mail: \email{zon53@mail.ru}}
\author{\speaker{Artem Popov}\\
        Department of Theoretical Physics, Moscow State University, 119992 Moscow, Russia\\
        E-mail: \email{ar.popov@physics.msu.ru}}
\author{Alexander Studenikin\\
        Department of Theoretical Physics, Moscow State University, 119992 Moscow, Russia\\
        Joint Institute for Nuclear Research, Dubna 141980, Moscow Region, Russia\\
        E-mail: \email{studenik@srd.sinp.msu.ru}}

\abstract{We develop the approach to the problem of neutrino oscillations in a magnetic field introduced in \cite{Popov:2019nkr} and extend it to the case of three neutrino generations. The theoretical framework suitable for computation of the Dirac neutrino spin, flavour and spin-flavour oscillations probabilities in a magnetic field is given. It is shown that there is an entanglement between neutrino flavour and spin oscillations
and in the general case it is not possible to consider these two types of neutrino oscillations separately. The closed analytic expressions for the probabilities of oscillations are obtained accounting for the normal and inverted hierarchies and the possible effect of CP violation. In particular, it is shown that the probabilities of the conversions without neutrino flavor change, i.e. $\nu_e^L \rightarrow \nu_e^L$ and $\nu_e^L \rightarrow \nu_e^R$, do not exhibit the dependence on the CP phase, while the other neutrino conversions are affected by the CP phase. In general, the neutrino oscillation probabilities exhibit quite a complicated interplay of oscillations on the magnetic $\mu_{\nu} B$ and vacuum frequencies. The obtained results are of interest in applications to neutrino oscillations under the influence of extreme astrophysical environments, for example peculiar to magnetars and supernovas, as well as in studying neutrino propagation in interstellar magnetic fields.}

\FullConference{
European Physical Society Conference on High Energy Physics - EPS-HEP2019 -\\
			10-17 July, 2019\\
			Ghent, Belgium}

\begin{document}

\section{Neutrino stationary states in a magnetic field}
To describe neutrino oscillations in an external magnetic field, we use the approach based on Dirac equation stationary solutions.
Here we assume that neutrino is a Dirac particle, and also possesses only diagonal magnetic moments. Then the stationary Dirac equations for each of the mass states are decoupled:
\begin{equation}\label{eq1}
(\gamma_{\mu} p^{\mu}-m_i-{\mu_i}{\bm{\Sigma}\bf{B}})\nu_i^s (p)=0,
\end{equation}
where $\mu_i$ are the neutrino magnetic moments, $i = {1,2,3}$ and $s = \pm 1$ is a spin number.

The above equation can be rewritten in the following form

\begin{equation}
\hat{H_i}\nu_i ^s= E \nu_i ^s,
\end{equation}
where the Hamiltonian is
\begin{equation}\label{Ham}
\hat{H_i} = \gamma_0 \bm{\gamma}\bm{p} + m_i \gamma_0 + \mu_i \gamma_0 \bm{\Sigma}\bf{B}.
\end{equation}

In order to classify the stationary solutions of eq. \ref{eq1}, we utilize the following spin operator
\begin{equation}\label{spin_oper}
\hat{S}_i = \frac{m_i}{\sqrt{m_i^2 B^2 + p^2 B^2_{\perp}}} \left[ \bm{\Sigma}{\bf B} - \frac{i}{m_i}\gamma_0 \gamma_5[\bm{\Sigma}\times\bm{p}]\bf{B}\right].
\end{equation}

It is easy to show that the spin operator commutes with the Hamiltonian \ref{Ham}, and its eigenvalues are $\pm 1$. Thus, we can use this operator to classify the stationary solutions:
\begin{equation}
\hat{S}_i \ket{\nu_i^s} = s \ket{\nu_i^s}, s = \pm 1,
\end{equation}
\begin{equation}\label{normalizaton}
\braket{\nu_i^s|\nu_k^{s'}} = \delta_{ik}\delta_{ss'}.
\end{equation}

For the neutrino energy spectrum in a magnetic field we have
\begin{equation}\label{spec}
E_i^s=\sqrt{m_i^2+p^2+{\mu_i}^2{B}^2 +2{\mu_i}s\sqrt{m_i^2{B}^2+p^2B_\bot^2}},
\end{equation}

This expression can be simplified if one accounts for the relativistic neutrino energies ($p \gg m  $) and also for realistic values of the neutrino magnetic moments and strengths of magnetic fields ($p \gg\mu B $). In this case we have
\begin{equation}\label{energy_s}
E^{s}_i \approx p + \frac{m^2_i}{2p} + \frac{\mu_i^2 B^2}{2p} + \mu_i s B_{\perp}.
\end{equation}

\section{Neutrino oscillations in a magnetic field probabilities}
In the most general form, neutrino oscillations probabilities can be represented in the following form
\begin{equation}
P(\nu_{\alpha}^h \rightarrow \nu_{\beta}^{h'}) = \left|\braket{\nu_{\beta}^{h'}(0)|\nu_{\alpha}^h(t)}\right|^2 = \Big| \sum_{i,k} U^*_{\beta i} U_{\alpha i} \braket{\nu_i^{h'}(0)|\nu_i^h(t)}\Big|^2,
\end{equation}
where $U_{ik}$ is the PMNS-matrix, $i,k = \{1,2,3\}$, $\alpha, \beta = \{e,\mu,\tau\}$ are flavours and $h,h' = \pm 1$ are helicities.
Thus, the amplitudes $\braket{\nu_i^{h'}(0)|\nu_i^h(t)}$ are the quantities of interest for us. Using quite the same approach based on the eigendecomposition as was implemented in \cite{Popov:2019nkr}, the amplitudes can be computed as follows
\begin{equation}
\braket{\nu_i^{h'}(t)|\nu_i^h(0)} = \sum_s C_{is}^{h'h} e^{-i E_i^s t},
\end{equation}
where the decomposition coefficients are
\begin{equation}\label{decomp_coeffs}
C_{is}^{h'h} = \bra{\nu_i^{h'}} \hat{P}_i^s \ket{\nu_i^h},
\end{equation}
and the projection operators were introduced
\begin{equation}\label{projectors}
\hat{P}_i^{\pm} = \ket{\nu_i^{\pm}}\bra{\nu_i^{\pm}} = \frac{1 \pm \hat{S}_i}{2}.
\end{equation}

The oscillations probabilities can be finally written as
\begin{equation}
P(\nu_{\alpha}^h \rightarrow \nu_{\beta}^{h'}) = \Big|\sum_s \sum_{i} U^*_{\beta i} U_{\alpha i} C_{is}^{h'h} e^{-i E_i^s t}\Big|^2,
\end{equation}
or in expanded form
\begin{eqnarray}\nonumber\label{prob_general}
P(\nu_{\alpha}^h \rightarrow \nu_{\beta}^{h'}) &=& \delta_{\alpha\beta}\delta_{h h'} - 4 \sum_{\{i,j,s,\sigma \}} \operatorname{Re} ([A_{\alpha\beta}^{h h'}]_{i,j,s,\sigma}) \sin^2\left(\frac{E_i^s-E_j^{\sigma}}{2}\right)t \\ &+& 2 \sum_{\{i,j,s,\sigma \}} \operatorname{Im} ([A_{\alpha\beta}^{h h'}]_{i,j,s,\sigma}) \sin\left(E_i^s-E_j^{\sigma}\right)t,
\end{eqnarray}
where the amplitude coefficients were introduced
\begin{equation}\label{ampl}
[A_{\alpha\beta}^{h h'}]_{i,j,s,\sigma} = U^*_{\beta i} U_{\alpha i} U_{\beta j} U_{\alpha j}^* C_{is}^{h'h} \left(C_{j\sigma}^{h'h}\right)^*,
\end{equation}
and
\begin{equation}
\sum_{\{i,j,s,\sigma \}} = \sum_{i>j;s,\sigma} + \sum_{s>\sigma;i=j}.
\end{equation}

As we can see, the probabilities of neutrino flavour, spin and spin-flavour oscillations in general case exhibit quite a complicated interplay of oscillations on different frequencies, that depend on both magnetic $\mu_i B_{\perp}$ and vacuum frequencies $\frac{\Delta m_{ik}^2}{4p}$. The same phenomenon in the two-flavour case was considered in \cite{Popov:2019nkr,Kurashvili:2017zab}. Throughout the derivation of the final expression \ref{prob_general}, we did not make any assumptions concerning the particular form of the mixing matrix $U$, except it being unitary. Thus, the general expression \ref{prob_general} can be used to calculate explicit expressions for the probabilities of neutrino flavour, spin and spin-flavour oscillation in a magnetic field in the three-flavour case. It is easy to show that the two-flavour analytical expressions obtained in \cite{Popov:2019nkr} can be retrieved from \ref{prob_general} by setting the mixing angles $\theta_{13} = \theta_{23} = 0$, i.e. decoupling $\nu_{\tau}$ from $\nu_e$ and $\nu_{\mu}$. Here we use the standard parametrization of the mixing matrix

\begin{equation}\label{U_PMNS}
U=\begin{pmatrix}
    1 & 0 & 0 \\
    0 & c_{23} & s_{23} \\
    0 & -s_{23} & c_{23}
  \end{pmatrix}
\begin{pmatrix}
    c_{13} & 0 & s_{13} e^{-i\delta} \\
    0 & 1 & 0 \\
    -s_{13} e^{i\delta} & 0 & c_{13}
  \end{pmatrix}
\begin{pmatrix}
    c_{12} & s_{12} & 0 \\
    -s_{12} & c_{12} & 0 \\
    0 & 0 & 1
  \end{pmatrix},
\end{equation}
where $\delta$ is the CP-violation phase for Dirac neutrinos. To describe antineutrino oscillations, we must replace $U$ with $U^*$.

Within the three-flavour case, it becomes possible to investigate the way non-zero Dirac CP-phase affects the probabilities of neutrino oscillations in an external magnetic field. Using the general expression \ref{prob_general}, it is possible to examine some features interesting analytically. For instance, it is easy to show that

\begin{equation}
P(\nu^L_{e} \rightarrow \nu^R_{e}) = P(\overline{\nu^L_{e}} \rightarrow \overline{\nu^R_{e}}),
\end{equation}
i.e. the survival probability of an electron left neutrino does not depend on the value of the CP-phase.

Moreover,

\begin{equation}
\sum_{\alpha} P(\nu^L_{e} \rightarrow \nu^R_{\alpha}) = \sum_{\alpha} P(\overline{\nu^L_{e}} \rightarrow \overline{\nu^R_{\alpha}})
\end{equation}
for $\alpha = e,\mu,\tau$. We conclude, that CP-violation cannot influence the total observed flux of right sterile neutrinos. However, the probabilities of conversions to $\nu_{\mu}^R$ and $\nu_{\tau}^R$ exhibit dependence on the CP-phase.

In general, it is quite hard to deal with analytical expressions due to their complicatedness. From now on we calculate \ref{prob_general} numerically.

Probabilities of neutrino flavour oscillations in a magnetic field $B = 10^{11} G$ for the cases $\delta = 0$ and $\delta = \frac{\pi}{2}$ is shown in Fig.1. Magnetic fields of such strength can be achieved inside astrophysical object, for example magnetars and supernovas. We also supposed that $\mu_1 = \mu_2 = 10^{-12} \mu_B$, which is consistent with the experimental upperbounds on neutrino magnetic moments, and neutrino energy is $p = 1$ MeV.

\begin{figure}[ht]
\begin{minipage}[h]{0.49\linewidth}
\center{\includegraphics[width=1\linewidth]{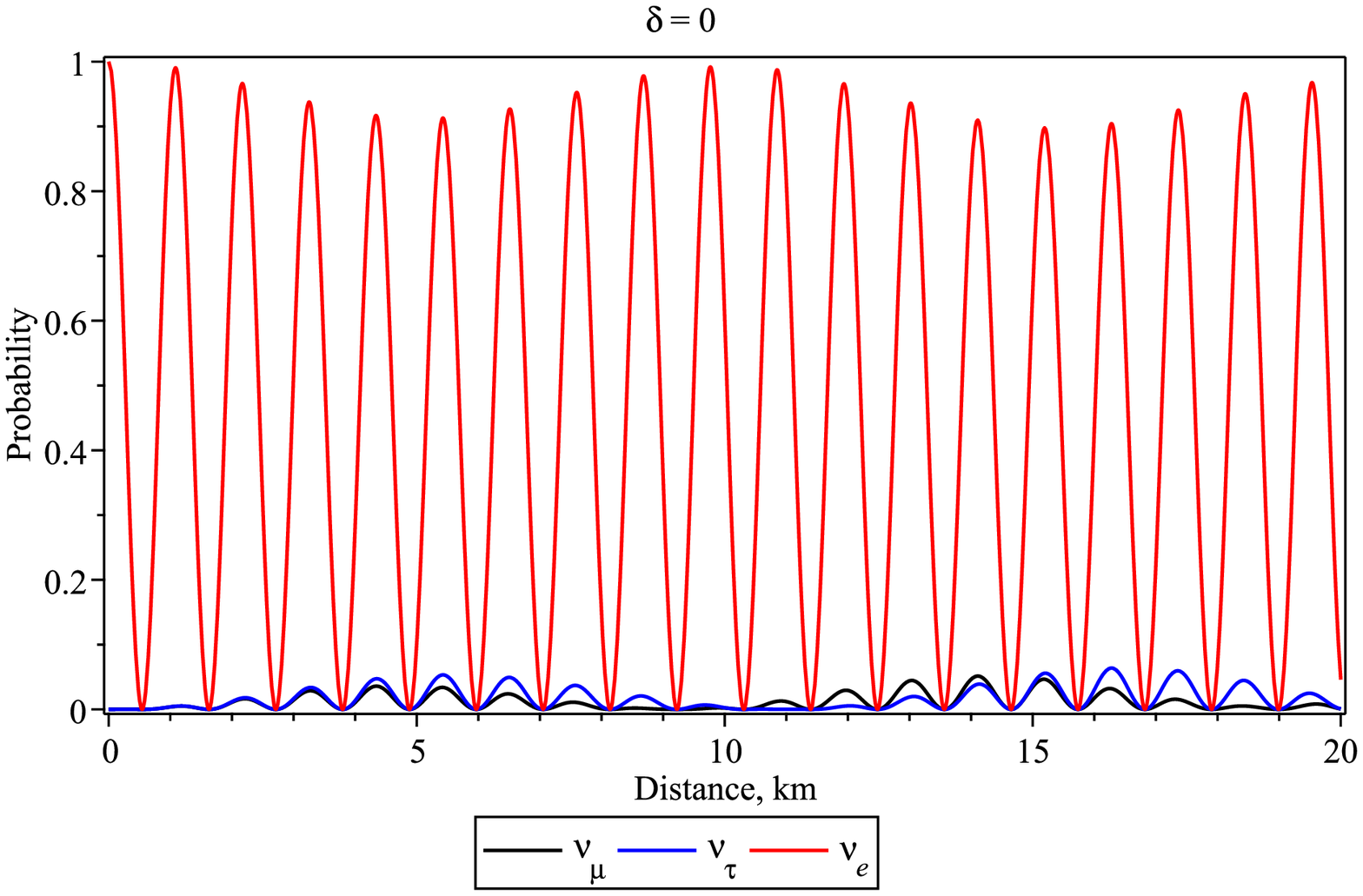}}
\end{minipage}
\begin{minipage}[h]{0.49\linewidth}
\center{\includegraphics[width=1\linewidth]{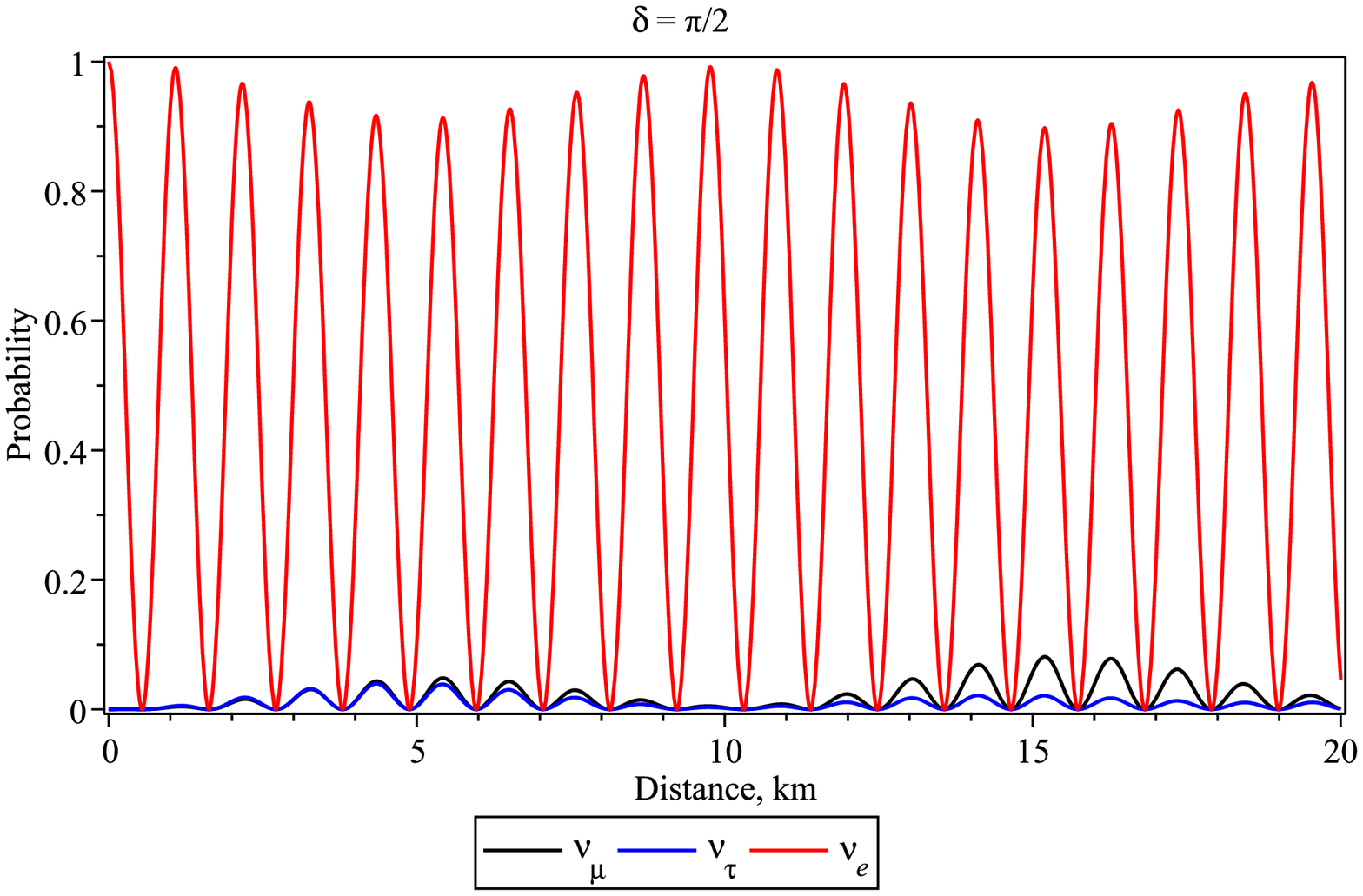}}
\end{minipage}
\caption{Probabilities of neutrino oscillations in a magnetic field $B=10^{11} G$ for the cases $\delta = 0$ and $\delta = \frac{\pi}{2}$.}
\end{figure}

Clearly, the electron left neutrino survival probability remains unaffected by the value of the CP-phase, as it was mentioned above. Meanwhile, the probabilities of oscillations to left muon neutrino and left tau neutrino has different phases.

\section{Conclusion}
We have generalized the approach to the problem of neutrino oscillation in an external magnetic field, developed in \cite{Popov:2019nkr}, for the case of three neutrino flavours. The obtained expressions can be reduced to those known before by setting $\theta_{13} = \theta_{23} = 0$. The effect of non-zero CP-phase is examined. It is shown that only $P(\nu_e^L \rightarrow \nu_{\mu}^L)$ and $P(\nu_e^L \rightarrow \nu_{\tau}^L)$ exhibit dependence on $\delta$. Using our approach, it is also possible to investigate the mass hierarchy effect on the phenomenon of neutrino oscillations in a magnetic field, but this is beyond the scope of the present paper. The obtained results are of interest in applications to neutrino oscillations under the influence of extreme astrophysical environments, for example peculiar to magnetars and supernovas, as well as in studying neutrino propagation in interstellar magnetic fields.

One of the authors (A.S.) is thankful to Michael Tytgat, the Chair of the Local Organizing Committee, and Barbara Clerbaux, the Vice-Chair, for the invitation to attend the European Physical Society Conference on High Energy Physics 2019 and for the excellent organization of the event.

\end{document}